\documentstyle[12pt,epsf]{article}

\newcommand{\gsim}{\lower.7ex\hbox{$\;\stackrel{\textstyle>}{\sim}\;$}}
\newcommand{\lsim}{\lower.7ex\hbox{$\;\stackrel{\textstyle<}{\sim}\;$}}

\def\beq{\begin{equation}}
\def\eeq{\end{equation}}
\def\bea{\begin{eqnarray}}
\def\eea{\end{eqnarray}}
\def\bq{\begin{quote}}
\def\eq{\end{quote}}

\def\bq{\begin{quote}}
\def\eq{\end{quote}}

\begin{document}

\baselineskip 24pt
\newcommand{\sheptitle}
{WHAT WILL WE LEARN IF A HIGGS BOSON IS FOUND?}

\newcommand{\shepauthor}
{G. L. Kane$^1$, S. F. King$^2$ and Lian-Tao Wang$^1$ }

\newcommand{\shepaddress}
{$^1$Randall Physics Laboratory, University of Michigan,
Ann Arbor, MI 48109-1120\\
$^2$Department of Physics and Astronomy, University of Southampton,
Southampton, SO17 1BJ, U.K.}

\newcommand{\shepabstract} {We examine what can be deduced from the
discovery of a Higgs boson,
focusing on a possible discovery at LEP. \ The analysis begins with the most
general situation where no further constraints can be deduced, and then
specializes to include various assumptions.
Assuming naturalness,
the relatively large mass suggests that one or
more of larger $\tan \beta$, large phase, or extra singlets is present.
We discuss the implications of 
a Higgs mass discovery for the SUSY spectrum,
and in particular gluinos and stops.
We show that a lighter gluino produceable at the Tevatron remains
a likely possibility, in disagreement with a recent claim.}

\begin{titlepage}
\begin{flushright}
hep-ph/0010312
\end{flushright}

\begin{center}
{\large{\bf \sheptitle}}
\bigskip \\ \shepauthor \\ \mbox{} \\ {\it \shepaddress} \\ \vspace{.5in}
{\bf Abstract} \bigskip \end{center} \setcounter{page}{0}
\shepabstract
\begin{flushleft}
\today
\end{flushleft}
\end{titlepage}

\section{Introduction}
\qquad The discovery of a Higgs boson will be of profound importance.
Arguably, it is one of the most important experimental results of all
time. \ The Higgs boson is a new kind of matter, the first kind to be found
in a
century. \ New quarks and leptons have been found, but they are all
particles like the electron that carry other charges. \ New force-mediating
bosons have been found but they are all quanta like the photon. \ The Higgs
boson is the quantum of a new kind of field, one for which the energy
density of the universe is lower when the field has a non-zero value, and
one whose interactions with other particles (both bosons and fermions) allow
them to have mass. The Higgs boson both completes the Standard Model of
particle physics, and points to how to extend the Standard Model. \ This
discovery will also be a remarkable achievement for human intelligence,
which was able to generate the idea of such a
particle in nature and then to construct experimental facilities in
which it could be
produced and detected. It will lead us into a new realm of understanding 
the laws of nature.

\bigskip

\qquad In view of the profound importance of the Higgs discovery
the present paper is designed to be largely accessible to the
non-specialist. The purpose of the paper is to
discuss what the measured values of mass and
production rate (assuming a discovery) can tell us about the parameters
of the Higgs sector, and
about how the Standard Model (SM) will be extended.
Many of the points in this paper will hold whether a Higgs boson is found at
LEP or at Fermilab. However,
we will focus on the possibility that the discovery 
occurs at LEP, and in
that case the mass is already approximately known (since the collider is
running at
or near its highest energy), and the cross section times branching
ratio must be as large as possible  to get enough events to
observe them. \ Initially, confirmation that a discovery is indeed a
Higgs boson will rely on seeing a decay dominated by couplings
proportional to mass, i.e., mainly to $b\bar{b}$, with a
$\tau \bar{\tau}$ branching ratio (at about $8 \%$) observed later, plus
observing the expected production rate approximately. One important
question is to identify whether the observed Higgs boson is that of the 
SM, a light Higgs boson of the
supersymmetric (SUSY) Higgs spectrum, or a composite state. 
If the mass of the Higgs boson is about 115 GeV,
one can argue that it is not the SM Higgs boson,
because it is well below the lower limit (of about 130 GeV) that
destabilizes the vacuum \cite{sherrev}.  \ Further, while in composite
approaches to the origin of mass one
cannot formulate theories well enough to make precise predictions, the
natural mass for a scalar in such theories is near the non-perturbative
limit. (For example, ref \cite{hamed00} gives an 
approximate region $165 < m_h < 230$ GeV, given certain assumptions.)
Thus it is very unlikely that a discovery at LEP could be interpreted as
a composite Higgs. 
In contrast there is motivation to study the case of a SUSY 
Higgs boson for several reasons.
It is well known that the mass of a light scalar is not 
protected against radiative corrections that raise it up
to the unification or Planck scale except in the presence of a symmetry
such as SUSY. In addition SUSY predicts a Higgs boson mass consistent
with a LEP measurement. Therefore we presume
that such a discovery would be the lightest Higgs boson of the
supersymmetric theory and proceed to study it in that context.
However if the
Higgs boson is indeed found at LEP the mass and cross section do not allow
us to prove from its properties that it is explicitly a supersymmetric
Higgs boson \cite{scalar}. 
Direct evidence for SUSY
will have to come from discovery of
superpartners or of some of the additional Higgs bosons
at LEP, Fermilab or LHC. Less direct evidence for SUSY could come
from observed deviations of the Higgs boson 
properties from the SM values when more
precise data is available.

\bigskip

\qquad It is known that in the general case the number of parameters needed
to describe the Higgs sector in SUSY models is at least seven
\cite{KB98, pilaftsis:1998, wagner:1999hb, KW00}. \
At tree
level the number is only two, but (as was first pointed out in 
Ref.\cite{KB98}) the rest enter via large top-stop loop
contributions to the Higgs potential that can shift the mass and cross
section and branching ratios significantly. \ If $\tan \beta $, the
ratio of Higgs vacuum expectation values, is large, or
if the theory is an extension of
the minimal supersymmetric SM (MSSM), there are
surely more parameters (from b-\~{b} loops, or from the mixing of additional
scalars). \ Other loop contributions such as W-chargino and
charged Higgs loops can be important as well
(see for example \cite{nath00} ) . 
If the mass and cross section are
measured there are two relations among the seven or more parameters, and
without further
input there are unfortunately no additional results that can be deduced,
apart from the general results of the above two paragraphs. 
It should also be noted that $115$ GeV is pushing
the constraints
on allowing electroweak baryogenesis in the MSSM
\cite{baryogen}, though those
limits are soft ones. 

\bigskip

\qquad Perhaps the most important success of supersymmetry
is that it provides a radiative mechanism by which the breaking of the
electroweak (SU(2)$\times $U(1)) symmetry can be explained. 
If radiative electroweak symmetry breaking is 
indeed correct, it leads to a relation between the measured W or Z masses
and the soft supersymmetry breaking masses. \ If this relation is indeed
physics, it must not involve fine-tuned differences of large numbers. \ If
we assume this is so then we get important constraints on the parameter
space. \ This is often called the naturalness assumption
\cite{ftorign, finetune}. \
This
assumption can be checked a posteriori since it implies the production
of
gluinos at the Tevatron, and has further implications that we describe below.

\bigskip

Given that a LEP Higgs boson suggests a state with full strength couplings
to W, Z and  small invisible branching ratio, we can usefully estimate the
number of Higgs bosons that will be produced at the Tevatron. Confirmation
of a signal in a given mass bin will require less data than a discovery.  
Once the Higgs
mass is measured (e.g. at LEP), all the Higgs bosons produced at the
Tevatron \cite{workshop} (even the inclusive channel) can
be studied to confirm the signal and 
to observe  smaller decay branching ratios, etc. The expected
number is
of order $15,000$ per $10$$fb^{-1}$. If the Tevatron collider and
detectors achieve design luminosity and are not limited by funding
shortages, they will gather $\sim 30fb^{-1}$. Thus with a known mass we can
confirm a signal and even 
hope to
see even $BR(\gamma \gamma)$ (which is expected to be $\sim 10^{-3}$.)
Observing the
$\gamma \gamma$ decay
immediately proves that $h$ does not have spin 1, and the value of the
branching ratio is a useful constraint on other new particles in the
theory.

\bigskip

\qquad The mass of 115 GeV is actually somewhat large for what is expected
in
the minimal theory if we insist it is not unnatural \cite{finetune,
barbieri00}.\ The full parameter space of the MSSM allows
masses up to about 130 GeV, but if the soft phases are zero the region
above about 105 GeV is
essentially excluded by naturalness unless $\tan \beta $ is rather
large, in which case masses up to about 115 GeV are allowed.
Thus the LEP mass
could be evidence that $\tan \beta $ is large. Alternatively, with general 
phases one can have $m_h=115$ GeV for smaller $\tan \beta$, so the
LEP mass could suggest large phases. \ Also, when an extra
SU(2) singlet scalar is added to the theory the upper limit on the mass for
a given $\tan \beta $ and naturalness increases about 30 GeV
\cite{KKW93, extrafam}, so the
larger mass
could also be evidence for this extension of the MSSM. Of course, a mixture
of these effects could be present, or even more general extensions
of the MSSM are possible.

\bigskip

When the full SUSY parameter space is present (including the phase
which enters the Higgs potential), one
might wonder if the production rate can be larger than in the SM. 
The production cross section can never increase for any parameter over the SM
one, but the BR$(b \bar{b})$ can be larger than the SM one for special
parameters \cite{suppressbr} if the branching ratios to other channels
are suppressed.
This can at most give about a $15 \%$ increase in cross section times
branching ratio. 

\bigskip

Another issue that needs to be studied is the possibility that the
``observed'' scalar is actually the second lightest one, the lightest being
largely decoupled. The arises over a significant region of the parameter
space of the theories with extra scalars \cite{extrafam}. This effect can
also happen in the MSSM in the  presence of phases (a suggestion to that
effect has
appeared in Ref \cite{wagner:1999hb} as this paper was being written.)

\bigskip

In the following, we first discuss the general situation and then specialize
to increasingly specific approaches to the Higgs sector. 

\bigskip


\bigskip

\section{General Case}

The most general SUSY model may involve an explicit Higgs mass $\mu$, plus a
rather general scalar sector, and involve
many complex parameters. In such a framework one can deduce 
no further implications  from the measurement of the Higgs mass and
production. With rotations one can essentially consider here the two usual
Higgs doublets $H_U$ and $H_D$ plus a single gauge singlet $N$ whose vacuum
expectation value is responsible for generating the $\mu$ term due to
the coupling $\lambda N H_UH_D$.
In a recent analysis (which did not include the effect of phases) one of us
showed that in the NMSSM the low 
$\tan \beta$ region is technically natural and permitted
a strong first order phase transition consistent with electroweak
baryogenesis \cite{NMSSM}. The reason for the improved naturalness
of the NMSSM may be simply understood since the tree-level
Higgs mass squared formula is modified from its MSSM value 
to $M_Z^2 (\cos^2 2\beta + C\sin^2 2\beta)$ 
where $C= 2\lambda^2/(g^2+g'^2)$ \cite{KKW93}. 
As we shall see later an increased tree-level
Higgs mass demands less from radiative corrections.
The extra Higgs singlet state may mix with or decouple from the
physical Higgs state, leading to possible interesting effects.

Specializing to the MSSM, with an explicit Higgs mass term $\mu H_UH_D$,
it was first recognized a couple of years ago that the Higgs
parameter space had been oversimplified \cite{KB98, pilaftsis:1998,
wagner:1999hb, KW00} and that some of the parameters
could be complex. \ This has now been studied in some detail
\cite{wagner:1999hb, KW00}, \cite{babu:1998cp}-\cite{gunion00},
and possible
effects are large. \ All the extra parameters enter from loop diagrams that
contribute significantly to the Higgs potential. \ The Higgs mass
eigenstates and their couplings to gauge bosons and fermions can be
modified. \ Any conclusion about the Higgs sector --- limits or implications
of discoveries --- that are made without including the full parameter space
are not justified and may be simply wrong \cite{KW00}. \ The top-stop loop
is presumably
the dominant one and including it one can list the minimum set of parameters
on which the Higgs potential (and hence Higgs mass and couplings) depends
as $M_L^2$, $M_R^2$, $|A_t|$, $|\mu |$, $\phi_\mu + \phi_{A_t}$, 
$B=m_3^3$, $\tan \beta$.
\ The first three
are the left and right handed stop soft breaking
square masses and the stop trilinear soft
mass parameter, then the Higgs mass parameter $\mu$ , followed by the
phase combination \ that arises as the sum of the
phase of $\mu $ and the phase of $A_{t}.$ \ $B$ is the Higgs mixing
parameter in the soft Lagrangian. \ Finally there is tan$\beta .$ 
\ The Higgs mass is directly measured kinematically if there is a
signal at LEP. \ Similarly, if there is a signal the cross section is about
equal to the SM one, which implies that the masses of the heavier Higgs
bosons are large and they effectively decouple; this eliminates one
combination of parameters.  \ Thus there are five
parameters left. \ If tan$\beta $ 
is large there is an important bottom-sbottom loop too, with more
parameters, and chargino and neutralino and charged Higgs loops could also
be non-negligible \cite{nath00}. \ In the latter cases there are too many
parameters and
it is impossible to learn anything beyond the direct data and the
implications outlined in the introduction. \ 

\bigskip

\qquad However, we think it is worthwhile to proceed with simplifying
assumptions below and see what can be deduced if we make some assumptions. \
While the assumptions may be right, they may not, and the results should not
be believed just because otherwise we cannot learn anything --- that is
``lamppost physics''. \ Rather, as we will see, at least to some extent the
assumptions can be checked a posteriori with more data. For all cases we
consider, we can point out what analyses are useful, but final studies
and
implications have to be done by experimenters who can include efficiencies
and errors in appropriate ways.

\bigskip

\section{Naturalness}
\qquad We think the safest assumption is that if nature is indeed
supersymmetric at the weak scale, and this provides the actual explanation
of EWSB, then this explanation will be natural, as discussed above. \
In that case we can study the general parameter space and find that it 
is indeed somewhat reduced in size. \ Some lower values of tan$\beta $
are excluded, but only up to tan$\beta $ of about 3-4. \ The phase remains
general, unfortunately, with any value allowed. \ The reduction of the
parameter space can be seen most clearly in the $\mu$ parameter, which has
an upper limit of about $300$ GeV. This limit is not very sensitive to
other parameters, and not too sensitive to the unknown phase. It has
considerable implications, particular for the neutralino and chargino
masses, for $b \rightarrow s\gamma$ branching ratio and CP asymmetry, and
for the LSP cold dark matter relic density and detectability.

In the context of naturalness one may wonder about the 
question of phases in SUSY. If all the phases are large,
there must be some cancellations responsible
for the smallness of the neutron and electron electric dipole
moments (EDMs) \cite{BGK}, and given our present understanding this
cancellation mechanism apparently violates naturalness.  
However the cancellations required are model dependent,
and in some theories this mechanism could be entirely natural.
Another possibility is that the first two
families of squarks and sleptons are very heavy, while the third family ones
are lighter; then the Higgs potential is unchanged, but the contribution to
EDMs is reduced, so the appeal to cancellations is reduced. In this case the
needed mass hierarchy would have to be explained. 
Alternatively there may be small SUSY phases for some presently unknown
theoretical reason. It is to this possibility that we now turn.

\section{Small Phases, Large $M_A$ and Larger $\tan \beta$.}

From now on we set the phase in the Higgs potential to
zero, so CP is conserved,
and also assume that $B$ is rather large (which physically means
a large CP-odd peudoscalar mass $M_A>>M_Z$) since it is known that in this
limit the Higgs production cross-section becomes maximal, consistent
with a LEP observation.
In this case the Higgs potential only depends on the four parameters
$M_L^2$, $M_R^2$, $M_{LR}^2=m_t(A_t-\mu \cot \beta)$, $\tan \beta$,
where $B$ has decoupled,  
and the phase has disappeared leaving only the relative sign of 
$A_t$ and $\mu$ appearing in the combination $M_{LR}^2$. 
The first three of these parameters completely
determine the physical stop masses $m_1$, $m_2$
and stop mixing angle $\theta_t$.
Another important question to address is the sign of $\mu$.
In the Higgs production process we are considering, only the relative sign
between $\mu$ and the trilinear coupling $A_t$ is relevant. We found that
for any set of parameter with one relative sign ('$+$' for example) that is
consistent with the
signal, there is another point in the parameter space with opposite sign
that is
allowed. Therefore, we cannot decide the relative sign if we take the
full parameter space into consideration.

Given these assumptions (including naturalness) a corollary is that  
$\tan \beta $ is larger, say 5-20, the upper limit chosen to avoid
the complications of the bottom quarks and squarks.
Naturalness prefers larger $\tan \beta$
for two independent reasons: the tree-level contribution
to the Higgs mass $M_Z \cos 2\beta$ is maximised;
and the top quark Yukawa coupling is reduced (see later).

Assuming larger $\tan \beta$ the tree-level Higgs mass is near $M_Z$.
Can one make any statements about the stop spectrum from the
condition that its radiative corrections shift the Higgs mass
from $M_Z$ to about 115 GeV, a difference of about 25 GeV?  
There are two dominant radiative corrections
from the stop sector,
a ``degenerate'' contribution depending on $\ln (m_1^2 m_2^2/m_t^4)$,
and a ``hierarchical'' contribution
depending on $\ln(m_2^2/m_1^2)$ and proportional to the combination
$\sin ^22\theta_t (m_2^2-m_1^2)/2m_t^2$.
It is possible to obtain the required radiative corrections
from either a degenerate heavy stop spectrum with a small mixing
angle, or a hierarchical stop spectrum with a large mixing
angle and a light stop squark. Degenerate stop squarks with large
mixing is also possible. There is no way to choose among them
without further information. Thus from the point of view of the
low energy theory it is difficult to learn much about the
stop spectrum from the Higgs mass, although once one stop mass
is measured the remaining stop mass and mixing angle
will be constrained.

\section{High Energy Input Parameters}

We now explicitly assume that the low energy parameters such as the
stop parameters and Higgs soft masses are determined from some
input parameters at a high energy scale, which for 
definiteness may be taken to be the unification scale $10^{16}$ GeV.
The relevant high energy input parameters are 
$M_i(0), m_Q(0), m_U(0), m_{H_U}(0), m_{H_D}(0), A(0)$ 
(the high scale gaugino soft masses, squark doublet soft mass, stop singlet
soft mass, Higgs soft masses, and trilinear soft mass, respectively),
together with 
sign($\mu$) and $\tan \beta$,
where $|\mu |$ and $B$ are fixed by the
low electroweak symmetry breaking conditions
(unlike the previous approach where the low energy  
Higgs soft masses were fixed by these conditions.)

The radiative symmetry breaking mechanism involves the high energy 
Higgs soft mass squared $m_{H_U}^2(0)$ being positive but the
low energy value $m_{H_U}^2(t)$, where $t=\ln Q$ and $Q$ is the low
energy $\overline{MS}$ scale, having been driven negative.
Naturalness requires that $|m_{H_U}^2(t)|$ 
should not be much larger than ${M_Z^2}$, and in order to prevent
this we require $m_{H_U}^2(t)$ not to be too negative.
As mentioned larger $\tan \beta$ implies a smaller top Yukawa coupling
which helps not to drive $m_{H_U}^2(t)$ too negative.
Another requirement is that the high energy gluino mass $M_3(0)$ 
not be too large, since large gluino mass drives the squark mass squareds
$m_Q^2(t), m_U^2(t)$ very positive, and the coupled renormalisation
group equations drive $m_{H_U}^2(t)$ very negative in response.
However $m_Q^2(t), m_U^2(t)$ are essentially just the stop parameters
$M_L^2$, $M_R^2$ and the 115 GeV Higgs mass requires at least
one of them to be quite large,
as discussed, so subtlety  is required to avoid some
violation of naive naturalness \cite{finetune}.

Implications for the
SUSY spectrum of a 114-115 GeV Higgs have recently been explored assuming the
minimal SUGRA relations $M_{1/2}=M_i(0)$ and 
$m_0=m_Q(0)=m_U(0)=m_{H_U}(0)=m_{H_D}(0)$
\cite{Ellis}, and it was concluded there that $M_{1/2}>250$ GeV.
However we note that these authors restrict themselves to small values
of $m_0\leq 200$ GeV where the cosmological relic density calculated with
their assumptions is in the desired range. In our view, the
implications of knowing $m_h$ for cold dark matter
can be summarized as follows. While we do expect the LSP to be the lightest
neutralino, and to be a significant portion of the cold dark matter, the
actual calculation of the relic density is known to be very sensitive to
a variety of assumptions. For example, $\Omega_{LSP} \sim 1/3$ can be the
result of calculations with lighter sleptons but heavier squarks once one
does not insist on a single scalar mass $m_0$. Or the presence of phases can
lead to $\Omega_{LSP} \sim 1/3$ \cite{relicphase} when it would not if the
phases were zero. Or most of the relic
density can be created in non-thermal processes \cite{nonthermal}. While the
actual value of $m_h$ does enter all calculations, we do not think it leads
to any significant increase in understanding the CDM relic density  at the
present time, because numerical results require too many major assumptions.
In particular we find the assumptions of minimal SUGRA and $m_0<200$ GeV
to be unnecessarily restrictive.

As an illustrative point, we find that for larger values of $\tan \beta$ 
and taking $m_0=500$ GeV with 
$A(0)=-2m_0$ (in our convention above) that the $M_{1/2}$ 
required for a 114-115 GeV Higgs is reduced
to about 190 GeV for positive $\mu$ (the Higgs measurement does not
constrain the sign of $\mu$), with a significantly lighter stop
mass of about 240 GeV and a chargino mass of about 130 GeV.
If in addition we allow a slight violation of scalar universality
by increasing $m_{H_U}(0)$ by 25\% relative to the other scalar masses then
$M_{1/2}$ is reduced still
further to about 180 GeV, with a reduced stop mass of about 160 GeV,
and a chargino mass about 140 GeV. Maintaining scalar universality
but relaxing gaugino mass universality with $M_2(0)=2M_3(0)$,
and taking $m_0=500$ GeV and $A(0)=-2m_0$ as before, we find
that $M_3(0)$ required for a $114-115$ GeV Higgs is reduced to 150 GeV with
a very light stop mass
of 80 GeV but a heavier chargino of mass 240 GeV.
The effects of non-universality 
can be understood as resulting from changes in the $\mu$ parameter
due to changes in $m_{H_U}^2(t)$ due to renormalisation group
effects.
Similar effects can be obtained for $m_0\sim 1$ TeV,
$A(0)>1$ TeV but with reduced naturalness.

Does naturalness favour a lighter gluino, possibly in the Tevatron 
range? Given our assumptions we require some large parameter in the
stop sector to increase the Higgs mass by 25 GeV due to radiative
corrections. The origin of this large parameter may come 
either directly at the high scale, by for instance choosing
$m_0\sim 500$ GeV, $A(0)\sim 1$ TeV, in which case 
$M_3(0)\sim 150$ GeV and the gluino can be lighter.
Or it could arise from having relatively small squark
soft parameters at the high scale, but having a large gluino mass
which increases the stop parameters through renormalisation
group running effects. Both possibilities seem equally 
likely from the naturalness point of view. The lighter
gluino case is testable at the Tevatron.
Of course in the more general case with the phase present
then the gluino can be lighter than in the zero phase case,
for a given Higgs mass and $\tan \beta$.
In general when the theory is extended to improve naturalness
a lighter gluino will be favoured.

\section{Conclusion}

We have discussed in general terms the implications of a possible
Higgs boson discovery. 
We first argued that a Higgs boson discovery
at LEP is evidence for a SUSY Higgs boson. 
The most general supersymmetry theory is too complicated
to draw any further conclusions from a Higgs boson mass measurement. 
We think it is worthwhile to make a series of assumptions and examine the
consequences at each stage.
We first considered the NMSSM (which extends the minimal supersymmetric
theory to include an additional scalar) where naturalness can be improved 
relative to the MSSM
for low $\tan \beta$ and electroweak baryogenesis looks more
promising. Turning to the MSSM, we emphasised that
the presence of phases can lead to $\tan \beta$ being lower than expected
for a given Higgs mass.
The BR$(b \bar{b})$ can be larger than the SM one for special
parameters \cite{suppressbr} if the branching ratios to other channels
are suppressed, allowing at most about a $15 \%$ increase in cross section
times branching ratio. 
We then imposed naturalness where an upper limit on $\mu$ could be set.
Setting phases to zero and assuming large $M_A$ reduces the number
of parameters considerably and with those assumptions naturalness probably
implies larger $\tan \beta$. The stop spectrum must satisfy the 
constraint of increasing the Higgs mass by about 25 GeV; it can do this
in several ways.
We then considered the high energy
input parameters and showed that with these assumptions some subtlety is
required to preserve naturalness. 
We concluded that a
lighter gluino produceable at the Tevatron is likely when the theory is
extended to improve naturalness,
and will coincide with a lighter stop squark, since both effects
arise from a larger trilinear soft mass.
This conclusion disagrees with a recent result due to the 
unnecessarily restrictive nature of the assumptions in that case.

Finally it is still necessary to consider the
possibility that LEP will only set a bound
on the Higgs mass, and not make a discovery,
in which case the true supersymmetric Higgs mass limit is well below the
LEP kinematic limit.
In this case the cross section is not
saturated, and the true limit has to be deduced from the full parameter
space. \ For the mass, the limit is no longer about 115 GeV, but is instead
about 20\% lower,
about 95 GeV \cite{KW00, wagner:1999hb}. No discovery does not yet imply a
conflict with naturalness, since in the general parameter space a given  
cross section can occur for a lower mass.

\vspace{0.5in}

\noindent{\Large \bf Acknowledgment}

G.K. thanks A.Schwimmer for a question. We are very grateful to L.Everett
for discussions and comments on the paper, and to S.Rigolin for a read of
the paper. S.K. is grateful to the Randall Physics Laboratory at
the University of Michigan and the Aspen Center
for Physics where some of this work was done, and to PPARC for the
support of a Senior Fellowship.

\end{document}